# Strain Localization and Percolation of Stable Structure in Amorphous Solids

Yunfeng Shi and Michael L. Falk

Department of Materials Science and Engineering, University of Michigan, Ann Arbor, MI 48109-2136 USA

Spontaneous strain localization occurs during mechanical tests of a model amorphous solid simulated using molecular dynamics. The degree of localization depends upon the extent of structural relaxation prior to mechanical testing. In the most rapidly quenched samples higher strain rates lead to increased localization, while the more gradually quenched samples exhibit the opposite strain rate dependence. This transition coincides with the k-core percolation of atoms with quasi-crystal-like short range order. The authors infer the existence of a related microstructural length scale.

Localization of strain occurs spontaneously in a variety of disordered materials including metallic glasses, amorphous polymers, granular media, foams and colloids [1-4]. Physically based theories for localization are critical for failure analysis in amorphous solids and the prediction of rheological response and structural stability of granular media, foams and colloids during processing and storage. Metallic glass is a particularly timely example of an emerging amorphous material in which the spontaneous localization of strain presents the dominant failure mode. Metallic glass can be produced by a variety of processing routes ranging from casting at relatively low cooling rates of a few Kelvin per second in the case of bulk glass formers [5,6] to splat quenching [7] and ion beam assisted deposition, an extremely energetic non-equilibrium process [8]. Particularly under uniaxial tension the tendency for metallic glass to localize strain often results in dramatic failure via fracture along a single shear band [9].

Current theoretical concepts regarding softening and localization in these materials have relied on the concepts of free volume [10] and shear transformation zones (STZs) [11,12]. In free volume theories [10,11,13] excess free volume generated during deformation lowers the viscosity of the glass resulting in softening and localization [13]. Closely related STZ theories focus on the micro-mechanisms of shear, positing that local thermally activated and/or mechanically induced transitions of STZs control deformation. These models have been used to understand shear softening and shear thinning [14], and to analyze data on the emergence of localization[15]. It has recently been shown, however, that the elastic interactions between STZs may result in chains that are neither localized nor extensive [16]. None of the current theories, however, can be used to directly model the nucleation and propagation of shear bands, and many of the parameters in the current theories cannot unambiguously be related to specific atomic scale structures or the thermo-mechanical history of the glass.

In developing theoretical models for such disordered materials Liu and Nagel have proposed that the temperature, shear rate and density can serve as intensive non-equilibrium thermodynamic parameters, and that a phase diagram in the three-dimensional space described by these parameters could serve to characterize the transition from the ergodic flowing state, to the jammed, non-ergodic, state [17]. In this context investigating strain localization is akin to asking whether the external application of shear can induce the nucleation of a first order transition. This leads to the question of the nature of the appropriate order parameter and the nature of the unjamming transition. Recent work [18] has linked the reverse transition to the onset of a percolating k-core cluster [19] in simulations of repulsive spheres.

In this work we explore the connection between the percolation of areas of quasi-crystal-like short-range order (SRO) and the development of strain localization. This choice is motivated by a number of recent observations of the role of quasi-crystalline phases and packings in stabilizing glassy metals. It has been proposed that a thermodynamically stable glassy state with quasi-crystal SRO may nucleate from the melt in Al-Fe-Si alloys [20]. Also a number of investigations have shown nano-scale crystallites and quasi-crystalline order in bulk metallic glass formers. The quasi-crystallites in Zr-based metallic glasses appear not only to stabilize the glassy state [21], but also to enhance the strength of the glass [22]. Furthermore, we have observed that amorphization of quasi-crystal-like SRO can provide the softening mechanism in simulated nano-indentation [23].

We have performed two-dimensional molecular dynamics simulations of a binary system that has been used extensively to study quasi-crystal and amorphous thermodynamic and mechanical properties and also has a crystalline ground state [12,23-26]. The system consists of 20,000 atoms. Each atom is either of species S (small) or L (large), and all atoms interact via a Lennard-Jones potential. The full details of the potential are presented elsewhere [24]. We take the reference length scale, $\sigma_{SL}$, to be the distance at which the potential of the interaction between the two species is zero and the reference energy scale, $\varepsilon_{SL}$, is chosen to be the interaction energy between the two species. We choose our composition such that the ratio between the species is $N_L:N_S = (1+\sqrt{5}):4$ to be consistent with other studies of this system. $T_g$ of this system is

TABLE 1 Summary of sample preparation

| Schedule Number | $T_{start}$ ($T_g$) | $T_{end}$ ($T_g$) | Quench Time ($10^3 t_0$) | Quench Rate ($10^{-7}$ $T_g/t_0$) | Number of Samples | PE per Atom ($\varepsilon_{SL}$) | % Short Range Ordering |
|---|---|---|---|---|---|---|---|
| 1 | 1.38 | 0.092 | *Inst.* * | *Inst.* | 6 | -2.3726±0.0006 | 34.2±0.1 |
| 2 | 1.23 | 0.092 | *Inst.* | *Inst.* | 6 | -2.3886±0.0008 | 37.3±0.2 |
| 3 | 1.15 | 0.092 | *Inst.* | *Inst.* | 6 | -2.3987±0.0007 | 40.0±0.2 |
| 4 | 1.08 | 0.092 | *Inst.* | *Inst.* | 6 | -2.4153±0.0007 | 43.7±0.2 |
| 5 | 1.08 | 0.092 | 1 | 9850 | 10 | -2.4398±0.0006 | 49.9±0.2 |
| 6 | 1.08 | 0.092 | 10 | 985 | 10 | -2.4670±0.0005 | 57.2±0.1 |
| 7 | 1.08 | 0.092 | 50 | 197 | 10 | -2.4877±0.0003 | 63.2±0.2 |
| 8 | 1.08 | 0.092 | 100 | 98.5 | 10 | -2.4980±0.0006 | 66.3±0.2 |
| 9 | 1.08 | 0.092 | 500 | 19.7 | 10 | -2.5209±0.0004 | 73.4±0.2 |
| 10 | 1.08 | 0.092 | 1000 | 9.85 | 10 | -2.5314±0.0005 | 76.4±0.2 |

* *Inst.* refers to rescaling the velocity followed by annealing over approximately 0.1 ns.

known to reside at approximately 0.325 $\varepsilon_{SL}/k$, where $k$ is the Boltzmann factor. All the particles have the same mass, $m_0$, which will be the reference mass scale. The reference time scale will therefore be $t_0 = \sigma_{SL}\sqrt{m_0/\varepsilon_{SL}}$. For a typical material $t_0 \approx 1$ ps, and $\sigma_{SL} \approx 3$ Å. Interactions are neglected beyond a cutoff set by the distance at which the interaction potential falls below a critical value, $U_{\alpha\beta}(r_{c,\alpha\beta}) = 0.0163$ $\varepsilon_{SL}$, such that $r_{c,SL}$ is 2.5 $\sigma_{SL}$.

The model system was simulated using a standard leapfrog integration scheme applied to the Newtonian equations of motion [12]. During quenching periodic boundary conditions were imposed, and the coupling to the external heat bath was modeled using a Nose-Hoover thermostat [27]. During the mechanical testing no thermostat was imposed, and the top and bottom boundaries were maintained as periodic boundaries while the left and right boundaries were free surfaces in order to mimic a uniaxial tensile test. The geometry utilized here differs from other investigations of shear localization [28,29] in that the direction of maximum shear does not correspond an a-periodic simulation cell boundary which would complicate the determination as to whether localization arises due to bulk structural changes or boundary conditions.

Exploring the rate dependence of localization is critical because plastic deformation inherently involves the interplay of thermally activated relaxation and mechanically induced transitions. The strain was imposed by rescaling the boundaries perpendicular to the loading axis and the y-positions by a small amount at each time step at nine constant strain rates ranging from $5\times10^{-4}$ $t_0^{-1}$ to $10^{-6}$ $t_0^{-1}$. This was performed using the molecular dynamics algorithm described in Ref. [12]. Since the strain rates imposed were significantly lower than the inverse of the time that sound wave needs to travel across the system, inertial effects were negligible.

Table 1 summarizes the sample preparations. The initial conditions were created by starting from supercooled liquids which are equilibrated for times well beyond the relaxation time of the liquid as determined by analyzing atomic mean-square-displacements. Subsequent to equilibration the temperature of the liquid was reduced to 9.2% of $T_g$ by cooling at constant volume. The most gradually quenched sample was cooled at an effective rate of $9.85\times10^{-7}$ $T_g/t_0$ corresponding to a quench over approximately 1 µs. Other samples were quenched at rates as much as 1000 times higher. In addition a number of samples were produced by quenching instantaneously from melts of different temperatures by rescaling the particle velocities and then allowing the system to age for 100 $t_0$, approximately 0.1 ns. The residual pressure was released by uniformly expanding the sample at a strain rate $10^{-6} t_0^{-1}$. Free surfaces were then introduced by eliminating periodicity along the surfaces parallel to the y-axis followed by a relaxation for 10000 $t_0$, approximately 10 ns. In each case we chose to characterize the structure of the resulting material by the potential energy per atom of the sample prior to the mechanical test. A number of independent samples were created following each quenching schedule to determine the effects of sample to sample variation. The differences in energy between the various samples is on the same order of magnitude, approximately 2 kJ/mol, as that measured in as-quenched and pre-annealed metallic glasses using differential scanning calorimetry (DSC) [30].

Fig. 1 shows average stress-strain curves for samples produced according to quenching schedule 1 (instantaneously quenched), 5 (quickly quenched) and 9 (slowly quenched). All stress-strain curves show an initially steady increase in stress with strain at small strains. Steady flow is achieved monotonically for quenching schedule 1 and 5, while for quenching schedule 9, the stress drops dramatically between 3% and 5 % strain. This drop corresponds to the propagation of shear band across the entire sample.

In order to quantify the degree of localization we define a quantity that we will call the deformation participation ratio (DPR). The DPR is the fraction of atoms that appear to undergo a deviatoric shear strain larger than the nominal deviatoric shear strain of the entire sample. The deviatoric shear strain is extracted from the atomic positions using the procedure for extracting a best fit strain [12]. In a homogeneously deforming sample the DPR should be approximately 0.5. However, during a highly localized deformation the DPR should become negligible, approaching the ratio of the shear band width to the system size. Fig. 2 shows the DPR at 5% strain as a function of the initial potential energy per atom for different quenching schedules. Samples with high initial potential energy per atom, corresponding to the instantaneously quenched samples, also exhibit higher DPR. Fig. 3 shows the spatial distribution of deviatoric strain for three samples produced by quenching schedule 1, 5 and 9, respectively.

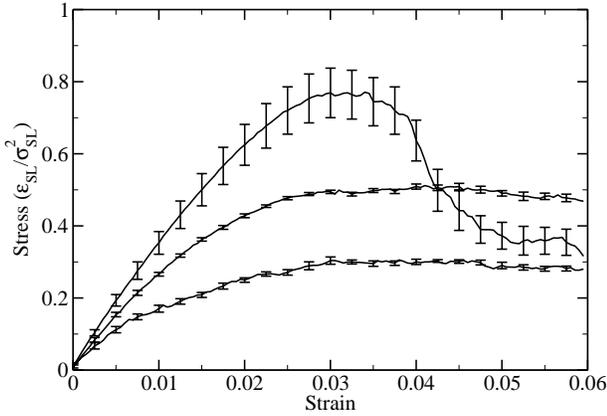

FIG. 1 Averaged stress-strain curves with error bars for, from bottom to top, instantaneously quenched samples (schedule 1), quickly quenched samples (schedule 5) and slowly quenched samples (schedule 9), respectively.

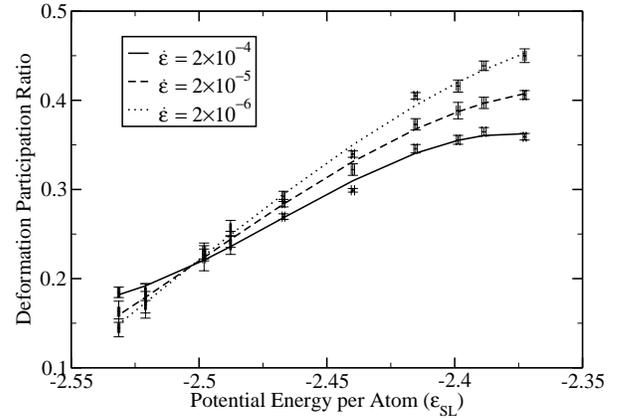

FIG. 2 DPR at 5% strain in uniaxial tension tests for strain rates of $2\times10^{-4}\ t_0^{-1}$, $2\times10^{-5}\ t_0^{-1}$, $2\times10^{-6}\ t_0^{-1}$ as function of the average potential energy per atom of the samples prior to testing. Each point represents a particular quenching schedule from Table 1.

The three lines on Fig. 2 differentiate simulations performed at three of the nine different strain rates simulated spanning more than two orders of magnitude. We utilize a power law relation to fit the strain rate dependence of the DPR, *i.e.* $DPR = A\dot{\varepsilon}^m$. Here $\dot{\varepsilon}$ is the strain rate, $A$ is a constant and $m$ is the strain rate sensitivity of the DPR. The power law form is not chosen to indicate any particular physical process and does not capture the fact that DPR saturates at 0.5, but this power law fit is useful to quantify the strain rate dependence. A non-negative $m$ indicates that in the limit of quasistatic loading and large system size the DPR approaches zero, corresponding to localized deformation; a negative $m$ indicates that the DPR approaches 0.5, corresponding to homogenous deformation. As shown in Fig. 4, at about $-2.507\pm0.003$ $\varepsilon_{SL}$ there is a change in sign of the strain rate sensitivity $m$, which corresponds to a discontinuous transition from localized flow to homogeneous flow in the quasistatic large N limit. The tendency toward distributed deformation at high strain rates in glasses prepared at lower quench rates is consistent with observations reported in Al, Pd and Mg based glasses [15,31].

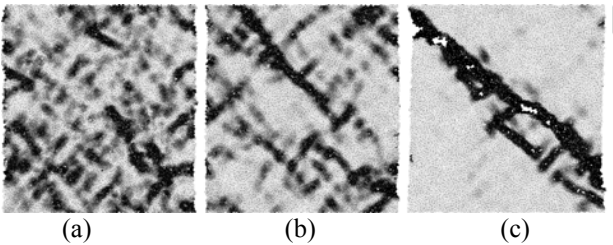

  (a)          (b)          (c)

FIG. 3 The spatial distribution of deviatoric strain in three samples produced by quenching schedules 1 (a), 5 (b) and 9 (c), loaded to 5% nominal strain at a strain rate of $10^{-6}$ ($1/t_0$). DPR for these three samples are 0.48, 0.36 and 0.20, respectively. Black corresponds to 10% strain, gray is 0% strain. The scale bar in the upper right of the image is 20 $\sigma_{SL} \approx 6$ nm.

Next we examined the dependence of the localization on the atomic scale structure of the samples. This system has an underlying quasi-crystalline state composed of nine local atomic motifs each consisting of an atom and its nearest neighbors [25]. We have analyzed the degree of SRO in the samples by determining if each atom resides in one of these motifs. The degree of SRO increases as the quenching time is extended. Low SRO regions are more likely to deform. For instance, for quenching schedule 5 these atoms are 25% more likely to deform. This implies that the percolation of the SRO could play an important role in determining whether localization nucleates or homogeneous deformation dominates.

We analyzed the percolation of the quasi-crystal-like SRO using both bond percolation and k-core percolation. In the latter each atom in the cluster must have at least $d+1$ = 3 neighbors also in the cluster, where $d = 2$ is the dimensionality [19]. Fig. 4 shows that the stable atoms form a bond percolating cluster when the initial potential energy per atom is $-2.464 \pm 0.007$ $\varepsilon_{SL}$ and they form a percolating k-core cluster at $-2.500\pm0.004$ $\varepsilon_{SL}$. This coincides closely with the initial potential energy per atom at which the strain rate sensitivity $m$ becomes non-negative.

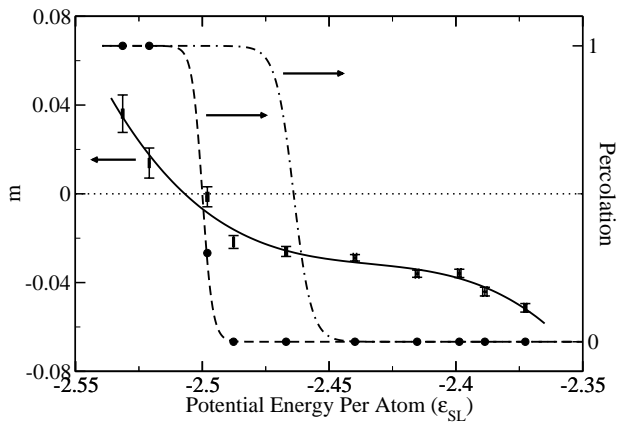

FIG. 4  Data points with error bars represent $m$, the strain rate sensitivity of DPR, as a function of potential energy of the samples prior to testing. The solid line is a fit to this data.  The other lines show the fraction of the specimens tested that exhibited k-core percolation of the SRO (dashed, circles) and bond percolation of the SRO (dot-dashed) as a function of potential energy.

In summary, we have performed a series of uniaxial tensile test simulations on binary models of a glass forming system in two dimensions.  More gradually quenched samples have a higher degree of quasi-crystal-like SRO. Upon the k-core percolation of SRO, the deformation mechanism changes from homogenous flow, where distributed deformation is favored at low strain rates, to localized deformation, where low strain rate favors shear localization. The degree of quasi-crystal-like SRO is in some respects the complement of the initial free volume in the free volume theory or the initial density of STZ defects in the STZ theory, but these simulation results imply that additional physics must be incorporated in these models to adequately capture localization.

In particular, incorporating the percolation of SRO requires that these theories include an order parameter that implicitly introduces a length scale.  Our best description of this order parameter, which we will refer to as the jamming length, would be the size of the largest k-core cluster that percolates through a coarse-grained region, or, if no such cluster exists, the largest k-core cluster contained entirely within the region.

The challenge for producing a theory of nucleation kinetics in such a system is that the ordered state in this case is not unique and may have a higher or lower connectivity of SRO depending on the thermo-mechanical history of the glass.  Put another way, these results suggest that, in analogy with the jamming phase diagram [17], the initial degree of SRO should be an additional axis on the deformation map originated by Spaepen [10] describing how temperature and strain rate/stress affect deformation behavior of metallic glasses.  While not a direct measurement of SRO, the degree of relaxation can be gauged experimentally by DSC studies of the glass prior to mechanical testing [30].

The authors would like to acknowledge the support of the NSF under grant DMR-0135009 and the UM Center for Advanced Computing under the NPACI program.